\documentstyle[twocolumn,fleqn,espcrc2,psfig,afterpage]{article}
\newcommand{\AmS}{{\protect\the\textfont2
  A\kern-.1667em\lower.5ex\hbox{M}\kern-.125emS}}

\def\lsi{\raise0.3ex\hbox{$<$\kern-0.75em\raise-1.1ex\hbox{$\sim$}}}
\def\gsi{\raise0.3ex\hbox{$>$\kern-0.75em\raise-1.1ex\hbox{$\sim$}}}

 {\begin{list}{}{%
  \setlength{\leftmargin}{10pt}%
  \setlength{\itemindent}{10pt}%
  \setlength{\rightmargin}{0pt}}}
  {\end{list}}

 {\begin{list}{}{%
  \setlength{\leftmargin}{8pt}%
  \setlength{\itemindent}{8pt}%
  \setlength{\rightmargin}{0pt}}}
  {\end{list}}

\begin{document}
\onecolumn
\title{Random manifolds and quantum gravity }

\author{{{A.~Krzywicki }}\address{Laboratoire de Physique
Th\'eorique, B\^at. 210, Universit\'e Paris-Sud, 91405 Orsay,
France$^{\rm 1}$
 }} 
       

\begin{abstract}

The non-perturbative,
lattice field theory approach towards the quantization
of Euclidean gravity is reviewed. Included is 
a tentative summary of the
most significant results and a presentation of the 
current state of
 art.
\end{abstract}

\maketitle  
\addtocounter{footnote}{1}
\footnotetext{ Unit\'e Mixte du CNRS UMR 8627. Orsay rep. LPT 99-51}
                 
\section{Preamble }

\noindent

The last year was
relatively quiet in this field. There are new results,  
interesting for experts, less for 
the audience of a plenary session. Therefore, 
I shall rather try to give you a tentative idea of what 
has been achieved since my plenary talk of 1988 and 
where we do stand now. For lack of space I quote only
the recent papers, omitting those quoted in the
monograph \cite{wein}. 

\section{Is Euclidean gravity worth attention?}

\noindent

The true subject of this talk is the statistical mechanics of
random manifolds, a subject of intrinsic interest. The 
partition function has the structure
\begin{equation}
Z = \sum_{ geometry} \; \;
\sum_{matter} e^{-action}
\label{z}
\end{equation}\noindent
The novelty is the summation over geometry, 
which is considered to be
dynamical. When the action depends 
only on the intrinsic geometry, the geometrical 
degrees of freedom are the topology and the metric of 
the manifolds. The summation over topologies being beyond 
control, one sums 
over all inequivalent metrics at fixed topology.
\par
 Euclidean
quantum gravity belongs to statistical mechannics.
Its relevance for the
gravity theory at $d>2$ is controversial. But it 
is, at least, an interesting theoretical 
laboratory, with its general covariance, perturbative 
non-renormalizability and bottomless action. 

\section{\bf The appeal of dynamical triangulations}

\noindent

The most promising discretization of Euclidean gravity 
rests on the idea of dynamical triangulations : the sum 
over geometries is defined to be the sum over all possible 
ways of gluing together equilateral simplices. 
\par
It is worth emphasizing the a priori beauty of this approach:
\par
- The arbitrariness of vertex labeling
replaces the reparametrization invariance in 
the continuum. 
\par
- The concept of a background metric never appears. 
\par
- No recourse is made to perturbation theory. 
Non-perturbative renormalization can be carried out 
using geometrical observables to set the scale. 
\par
- On a lattice  
the action is not bottomless. But for $d>2$
the most probable manifolds 
do {\em not} correspond to the bottom of the action. 

\section{The glory in two dimensions}

\noindent
 
In 2d the discrete model of pure gravity
can be solved exactly. The results match those obtained in
the continuum\footnote{The connection 
between the dynamical triangulations
and the Liouville theory is further 
studied in \cite{catmot}, where
the authors attempt to reconstruct 
the Liouville field in 
the discrete framework.}. 
Actually, the analytic power of the discrete theory 
often exceeds that of the continuum formalism.
The discussion can in many cases be extended to 2d gravity 
coupled to conformal matter fields. Let me briefly 
sample some highlights:
\par
- In 2d 
one has
\begin{equation}
\# triangulations(A)  \sim A^{\gamma - 3} e^{\kappa_c A} \; ,
\label{entr}
\end{equation}\noindent
where $A$ is the total area and $\gamma$ is 
the string susceptibility exponent.
 Summing over triangulations
with two marked points separated by the geodesic distance 
$r$ defines an invariant "two-point" function $G(r)$, whose 
explicit calculation for pure gravity is a great
success of the theory.
The scaling properties of $G(r)$ determine the two basic 
critical exponents: the Hausdorff dimension $d_H$ of 
space-time, which turns out to be $d_H=4$ (not the 
naive 2!) and $\gamma=-1/2$.
\par
- An exact solution has been found 
\cite{kazak} for the discrete $R^2$ gravity
in 2d (here $R$ denotes the scalar curvature). It has 
been proved that the infrared behavior of the system is 
that of the standard Liouville gravity for any finite 
$R^2$ coupling\footnote{However, is is well known 
that the surfaces become crumpled, when this coupling 
is large negative. The transition Liouville
$\rightarrow$ crumpled seems to be a cross-over.}. The
$R^2$ term flattens the surfaces locally, but at large 
scales they always look alike. 
\par   
- The equivalence between dynamical 
triangulations and matrix models can be used
to derive a number of results \cite{zinn}. E.g. the 
string susceptibility exponent can be calculated for all
unitary models where 2d gravity is coupled to conformal 
matter fields \footnote{The result was
first found for some discrete models, then generalized 
in the continuum framework and eventually rederived 
using matrix models.}  : writing $c=1-6/m(m+1)$ 
one finds for spherical topology
$\gamma = -1/m$. Of course, this holds at 
the critical point \footnote{The geometrical scaling 
at $c \neq 0$ is not yet understood.
Numerical simulations yield $d_H \approx 4$ for 
$0 \leq c \leq 1$. See also \cite{bow}. }. 
The topological expansion enables one to extend 
this result to higher genera \footnote{Attempts 
to go beyond the perturbative approach, 
via the so-called double scaling limit and the associated 
differential equations \cite{zinn}, did not provide, 
as yet, the expected insight into the non-perturbative 
physics of fluctuating topologies.}.
\par
 - The "gravitational dressing" of the scaling exponents
is under control. It depends on $c$ only\footnote{ A 
discussion of scaling exponents observed on quenched 
geometries thermalized with a "wrong" value of $c$ 
can be found in \cite{deslast}.}. 
The celebrated example is that of the Ising model on
a randomly triangulated manifold : the critical exponents 
have been calculated exactly and differ from the 
classical Onsager ones! 
\par
- It was long unclear whether there 
is any intermediate phase  between 
Liouville gravity and branched polymers, 
above the $c=1$ barrier. The answer seems to 
be found by in \cite{davc1}, using a renormalization 
group argument: the critical behavior at $c>1$ 
is generically that of branched polymers, but  
finite size effects are exponentially 
enhanced when $c \rightarrow 1$. 
Furthermore, David's argument  
suggested the existence of phenomena, which 
have eventually been found 
numerically \cite{begud}.
\par
The aim of these examples is to illustrate the claim that
2d "gravity" is at present the best, if not the only 
example of fully fledged quantum geometry. Other 
approaches to quantum gravity, including 
the most advertized ones, not only do not tell us 
much about the microscopic geometry of space-time, they 
do not offer yet a suitable framework to ask many 
relevant questions.
        
\section{\bf The boom of baby universes}

\noindent

One of the most interesting results in the field is the 
discovery that it is very unlikely for a generic random 
manifold to make small fluctuations around some more
or less smooth average configuration. If one lets it 
fluctuate freely, there are bubbles, called baby universes, 
growing out if it. Further, there are baby universes growing 
on baby universes an so on. The final structure is a fractal. 
\par
Using combinatorial arguments and (\ref{entr}) one can 
estimate the average number of baby universes with a 
given volume $B$ in a manifold of volume $A$ to be~:
\begin{equation}
n(B,A) \sim A [(1 - \frac{B}{A}) B)]^{\gamma - 2} 
\;,\; B < A/2 \; \; ,
\label{bu}
\end{equation}\noindent
so that $\gamma$ also controls the fractality of the 
manifolds \footnote{This result also holds for $d > 2$
provided (\ref{entr}) is true, with $A$ being the volume.}. 
From (\ref{bu}) one finds easily that the 
number of baby universes carrying a {\em finite}
fraction of the total volume behaves like $A^\gamma$. 
Clearly,  dramatic things happen when 
$\gamma > 0$: large baby universes proliferate and 
the manifolds degenerate into polymer-like structures.
\par
Recently, Ambj\o rn, Loll and collaborators \cite{loll}
have devised an alternative, model of 2d quantum gravity,
where the creation of baby universes does not occur. 
The consequence is that most of
the results listed in sect. 4 are gone:
there is no anomalous scaling, 
$d_H=2$, Ising model critical exponents
take Onsager's values etc. 
\par
Introducing a chemical
potential for minimal neck baby universes and using it 
to enhance the weight of configurations with large 
number of babies one produces an ensemble whose 
string susceptibility exponent is
that of branched polymers \cite{begud}.
\par
One could multiply such examples. They suggest that the dynamics 
of the Euclidean quantum gravity is to large extent controlled 
by the dynamics of the baby universes. Anticipating on the
discussion to follow let me remark that one of basic problems 
at $d>2$ is that we do not know how to tame the process of 
baby universe creation.

 \section{The improving tools}
 
\noindent

Reporting about the progress made in this field one should
mention the impressive development of the tools employed.
In short, people interested in random geometries 
have now at their
disposal the complete 
toolbox of a perfect lattice theorist:
local and ergodic algorithms,
a nonlocal algorithm \cite{surgery}, renormalization group
techniques, especially the node decimation one 
\cite{renor2}, long strong coupling series,
also for $d>2$ \cite{ourlat}.

\section{$d>2$ : What are telling the computers~?}

\noindent

\subsection{Discrete Einstein-Hilbert action}
Our understanding of $d>2$ rests on numerical simulations.
For $d=3$ it was rapidly established that the pure
gravity model has two phases, separated by a first order 
transition. In one phase the manifolds are elongated, 
resembling branched polymers, in the other they are crumpled.
A similar transition was found in 4d, but data were
compatible with it being a continuous one. Precise 
simulations at large volumes have shown that this 
transition is, in fact, of first order too \cite{bbkp}. 
This was confirmed by other studies \cite{otherd4}.
This year 4d simulations were carried out with 
the so-called degenerate triangulations. In \cite{bilgu} 
the double peak in the "energy" histogram is already 
seen at $N_4=4000$, while in \cite{bbkp} 
it is only observed at $N_4=32000$. This opens 
the possibility of studying the
effect in a wide volume interval.
\par
Incidentally, the data of \cite{bilgu} also strengthen the 
numerical evidence that the number of triangulations is in 4d
exponentially bounded. This bound, for $d>2$, is our community's 
contribution to topology!
\par
The dynamics of the transition is by now elucidated
\cite{hot}-\cite{cat2}. At finite volume the transition
occurs in two steps : first some singular vertices are formed,
then these vertices condense and in the crumpled phase one
finds a sub-singular link connecting two highly singular
vertices \footnote{The structure of the crumpled phase 
described here is not universal. With degenerate
triangulations one rather observes in the crumpled phase 
a gas of sub-singular vertices \cite{bilgu}.}.
\subsection{Models with modified action}
Attempts were made to soften the transition by modifying the 
action. The simplest such modification follows the old 
proposal by Brugmann and Marinari \cite{brma}: one weights the 
triangulations with 
\begin{equation}
weight \; factor = \prod_{j=1}^{N_0} o(v_j)^\alpha  \; ,
\label{bruma}  
\end{equation}\noindent
where $o(v_j)$ is the order of the vertex $v_j$ . 
Alternatively one can put the triangle order instead, this 
does not make much difference. Another modification, which at
the end of the day turned out to give very similar results, is
discussed in sect. 10.
\par
The phase diagram was already discussed at length  by
Thorleifsson last year. At sufficiently negative values of
$\alpha$ a new phase, baptized "crinkled" in \cite{suis},
appears. It looks smoother then the crumpled one, since the
Hausdorff dimension and $\gamma$ are finite and
$\gamma < 0$ . However, it seems that with increasing
volume the transition crumpled $\rightarrow$ crinkled 
runs towards large values of the Einstein term coupling, 
where $\langle N_0\rangle/N_4 \rightarrow
1/4$. This corresponds to the kinematic boundary. The
crinkled phase is also infested with sub-singular vertices.
\par
The dynamical triangulations near the kinematic boundary were
studied this year in 3d, using both the strong coupling series 
and the Monte Carlo simulations \cite{blast}. The transition 
branched $\rightarrow$ crinkled looks 
continuous, perhaps of third order. In the crinkled phase 
$\gamma < 0$ , but $d_H =\infty$ or $\approx 2$, 
depending on whether it is
measured on the triangulation or its dual, a feature hardly
compatible with the existence of a sensible thermodynamic 
limit.

\section{ $d>2$ : The same story by backgammon players}

\noindent

The structure of the phase diagram can be qualitatively 
understood within an exactly solvable model inspired by the 
backgammon game \cite{bbjlast}. The idea is to 
replace the sum over triangulations by the sum
over weighted partitions of vertex orders, the weight being 
$\propto o(v)^\alpha$. Set $r=N_0/N_4$. 
In the thermodynamical limit  the 
model has generically two phases : for $\kappa < \kappa_c$
one has $r=0$ and a singular vertex with order
$\sim N_4$, while for $\kappa > \kappa_c$ the value of $r$
is finite and all vertex orders are bounded. This 
mocks the transition crumpled $\to$ branched.
\par
For large enough negative $\alpha$ the system jumps at  
$\kappa = \kappa_c$  from $r=0$ to $r=1/4$,
and the order of the most singular vertex drops suddenly
(but remains $\sim N_4$). This is similar to the transition
crumpled $\to$ crinkled \footnote{
Strictly speaking, for $-2 < \alpha < -1$ 
the transition crumpled $\to$ branched becomes
continuous. This prediction has not really been checked. But
it is unlikely to be verified. The predictions of
the model should not be taken too literally. E.g.
simulations with degenerate triangulations in 3d indicate 
that in the crumpled phase $\langle N_0 \rangle /N_3$ 
approaches a {\em finite value}
$\approx 0.02$ when $N_3 \to \infty$
\cite{gudpriv}. Also, in 2d, the model does not
see the $c=1$ barrier}.

\section{ A generic instability?}

\noindent

No phase identified at $d>2$ is a serious candidate for a
physical space-time. The alternative, crumpled or branched, 
also occurs in 2d at $c>1$ \footnote{
The Liouville phase appears to be an 
exception. This can perhaps be understood:  the
system gravity+matter with $c<1$ seems
over-constrained (for a recent
discussion see \cite{catmot}).}. 
I am tempted to formulate the following conjecture:
{\em the generic random manifolds are unstable}.
If true, this would be a very interesting 
 result in 
statistical mechanics.  Somewhat
frustrating viewed from the quantum gravity perspective, 
although it cannot be excluded that the instabilities are a
lattice artifact and that the physical space-time is
recovered as one approaches some as yet undiscovered 
fixed point. My feeling is that we are rather missing 
some important part of the puzzle. It is a big 
challenge to discover it. Or,
if this were the case, to prove that the
constructive, lattice approach is inadequate. 

\section{The miseries of a brilliant idea}

One would like to control the dynamics of 
baby universes. In 2d and for $c>1$, 
there occurs a condensation of
metric singularities ("spikes")  
which can be avoided by
moving $c$ below unity. 
In \cite{jur} it was suggested that
a similar phenomenon might occur in 4d. But, in 4d
the sign of matter contributions 
to the conformal factor effective action is opposite
to that found in 2d and one can hope stabilizing the 
manifolds by {\em adding } conformal matter fields.
One should mention, however, that in 4d the idea 
is less founded than in 2d.   
\par
A numerical experiment testing that is
described in \cite{hope}, reporting
encouraging results. But it 
was found later, in \cite{suis}, that the model
with extra matter fields and that with measure 
modified \`a la Brugmann-Marinari are essentially 
equivalent. Hence, the absence of polymerization 
seen in \cite{hope} has here an
origin very different from the expected one.
\par
Thus the idea has failed. But the way it 
failed is interesting. The introduction of 
gauge matter fields produces a {\em local}
modification of the action. The simplest explanation 
of this fact is that the correlations between 
field fluctuations become short ranged. If so, 
the gauge fields have no chance 
to do the job they were supposed to do.

\section{Wishful thoughts about the future}

\noindent

Let me just mention 
briefly a few ideas which start receiving attention.
\par
In standard dynamical triangulations one sums over 
{\em all} simplicial complexes. Perhaps 
one should restrict the support of 
the Feynman integral. Can one figure out what 
a relevant constraint could be? 
This question is the starting point of \cite{loll}. 
They start in the Lorentzian regime in 2d 
and consider triangulations endowed 
with a causal structure. They eventually go to 
the Euclidean formulation, but with a dramatically 
reduced class of admissible triangulations. 
Baby universes are prohibited and the manifolds
are much smoother than in Liouville gravity. Actually,
they look too smooth, one would like quantum
gravity to me more entertaining, but perhaps this 
will be the case for $d>2$. 
\par
Another almost unexplored avenue is SUSY. It is essential
in string theory. Perhaps it is a 
necessary ingredient of any sensible
quantum gravity theory?
Perhaps with SUSY one can avoid the
localization of matter fields and revive the idea 
reviewed in sect. 10 \footnote{There exists 
a class of one-dimensional solvable models 
of disorder, where SYSY and delocalization are related.
Could this be a hint? These models are admittedly
very special.}. Another hint: according to 
\cite{kpz2} the $c=1$ barrier disappears
in $N=2$ world-sheet SUSY models.
One immediate problem: what SUSY, target 
space \footnote{An attempt in this direction is being
made in  \cite{oda}} or world sheet\footnote{A small step
in this direction is made in \cite{bjk}, where fermions
are put on a random manifold.}? 
The latter is necessarily broken on the lattice.
Perhaps too strongly broken to have significant 
consequences? But would the target
space SUSY be enough? Implementing SUSY is a
notoriously difficult problem of fundamental 
importance. Models in 2d should 
help developing intuition on 
the back-reaction of SUSY on geometry.
\par
In conclusion, much has been achieved,
but further surprises are not expected if 
one dooes not go beyond what 
became the common lore. We must be
"bold, yet bolder, even most bold". This is what Danton
said in 1792. One year later ...

\end{document}